\documentclass[amssym,useAMS,usenatbib,times,letters]{mn2e}

\usepackage{graphicx}
\usepackage{subfigure}
\usepackage{amssymb}

\usepackage[ascii]{inputenc}
\usepackage[T1]{fontenc}

\usepackage{amsmath,amssymb,amsfonts,textcomp}
\usepackage{color}
\usepackage{calc}

\definecolor{Blue}{rgb}{0,0.08,0.65}
\definecolor{Red}{rgb}{0.65,0.08,0.05}
\definecolor{Green}{rgb}{0.15,0.45,0.25}

\begin{document}

\title[The dusty, albeit ultraviolet bright infancy of galaxies]{
The dusty, albeit ultraviolet bright infancy of galaxies}

\author[J. Devriendt, C. Rimes, C. Pichon, R. Teyssier, D. Le Borgne et al.]{
J. Devriendt,\textsuperscript{1,2} C.
Rimes,\textsuperscript{2} C. Pichon,\textsuperscript{1,3}
R. Teyssier,\textsuperscript{4,6} D. Le
Borgne,\textsuperscript{3}    D.  Aubert,\textsuperscript{5} \newauthor E.
Audit,\textsuperscript{4} S. Colombi,\textsuperscript{3}
S Courty,\textsuperscript{2} Y.
Dubois,\textsuperscript{1}    S. Prunet,\textsuperscript{3}    Y.
Rasera,\textsuperscript{8}  A. Slyz,\textsuperscript{1}  D.
Tweed\textsuperscript{3 } \\
\textsuperscript{1}Astrophysics,
University of Oxford, Keble Road, Oxford OX1 3RH, UK. \\
\textsuperscript{2}Centre de Recherche Astrophysique de Lyon, UMR 5574,
 9 Avenue Charles Andr\'e, F69561 Saint Genis Laval, France. \\
\textsuperscript{3}Institut d'Astrophysique de Paris, UMR 7095, 
98\textsuperscript{bis} Boulevard
Arago, F75014 Paris, France.
\\
\textsuperscript{4}Service
d'Astrophysique, IRFU, CEA Saclay, Bat 141, F91191 GifsurYvette,
France.\\
\textsuperscript{5}Observatoire Astronomique, Universit\'e de
Strasbourg, 11 rue de l'Universit\'e, F67000 Strasbourg, France.\\
\textsuperscript{6}Institute of Theoretical Physics, University of
Zurich, Winterhurerstrasse 190, CH8057 Zurich, Switzerland.\\
\textsuperscript{7}Jeremiah Horrocks Institute for Astrophysics \&
Supercomputing, University of Central Lancashire, Preston PR1 2HE, UK. \\
\textsuperscript{8}Laboratoire Univers et Th\'eories, UMR 8102,
  5 Place Jules
Janssen, 92190 Meudon, France}

\maketitle


\begin{abstract}
{ The largest galaxies acquire their mass early on, when the Universe is
still youthful. Cold streams violently feed these young
galaxies a vast amount of fresh gas, resulting in very efficient star
formation. Using a well resolved hydrodynamical
simulation of galaxy formation, we demonstrate that these mammoth
galaxies are already in place a couple of billion years after the Big
Bang. Contrary to local starforming galaxies, where dust re-emits
a large part of the stellar ultraviolet (UV) light at infrared
and sub-millimetre wavelengths, our
self-consistent modelling of dust extinction predicts that a
substantial fraction of UV photons should escape from primordial
galaxies. Such a model allows us to compute reliably the number of high
redshift objects as a function of luminosity, and yields galaxies whose
UV luminosities closely match those measured in the deepest
observational surveys available. This agreement is remarkably good
considering our admittedly still simple modelling of the interstellar
medium (ISM) physics. The luminosity functions (LF) of virtual UV
luminous galaxies coincide with the existing data over the whole
redshift range from 4 to 7, provided cosmological parameters are set to
their currently favoured values. Despite their considerable emission at
short wavelengths, we anticipate that the counterparts of the
brightest UV galaxies will be detected by future sub-millimetre
facilities like ALMA.}
\end{abstract}

\section{Introduction}
{Over the past decade, the cold dark matter
model, complemented with dark energy ($\Lambda$CDM), has established itself
as the theoretical framework of choice to describe the formation of
structures in the Universe. Whilst dark matter dominates the dynamics of 
structure formation on large scales, a host of observational evidence 
indicates that the situation is reversed on galactic scales \cite{ref6}. 
In particular, the inner structure of galactic dark matter halos must be 
affected by baryonic processes \cite{ref7}. Whereas it is
certainly true that critical aspects of the baryonic physics of galaxy
formation are still poorly understood and beyond the reach of direct numerical
simulations \cite{ref5}, the overwhelming
majority of observational constraints unfortunately comes from
the electromagnetic emission of these baryons. Therefore, 
significant progress in our understanding of the
theory of galaxy formation and evolution depends on our ability to
perform hydrodynamical cosmological simulations. At first, these
will necessarily be complemented by appropriate subgrid modelling,
presumably based on knowledge gained from hydrodynamical simulations
performed on smaller scales \cite{ref8}. Indeed, we
are seeking to describe highly nonlinear mechanisms that are coupled
together, such as radiative gas cooling, star formation and feedback.
Such complexity makes it, for instance, very unlikely that even when
armed with a satisfying model for supernovae explosions, one can
describe analytically their interplay with a clumpy, multiphase
ISM and/or intergalactic medium (IGM).

\begin{figure}
\center  \includegraphics[angle=0,width=8.5cm]{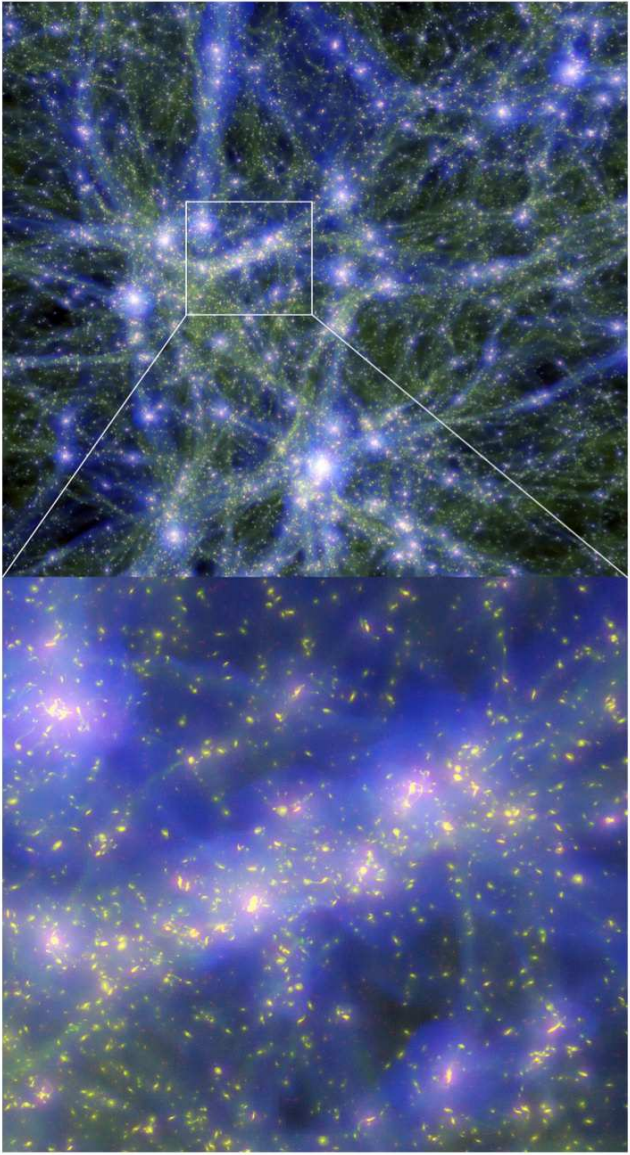}
\caption{
A global composite multiscale view of the Mare
Nostrum simulation described in the text. The blue colour traces the
gas temperature, green represents the gas density and red shows the
dark matter density. The box size is 50h$^{-1}$ co-Mpc
on a side for the top panel and 10h$^{-1}$ comoving Mpc for the bottom one.
 Galaxies appear as pinkish yellow smallscale disk-like structures.}
\end{figure}

Perhaps the most significant flaw of all galaxy formation models
which rely on the results of pure dark matter simulations is their neglect of
the filamentary cold gas flows which feed galaxies \cite{ref3, ref2}. Gas accreted 
in this mode substantially dominates the accretion budget. 
Indeed, cold flows are the unique mode of gas accretion for 
galaxies hosted by {\em all} dark matter halos with mass less than 
5 10$^{11}$ M$_{\sun}$ (i.e. the vast majority of galaxy size halos).
They are reported in every high resolution cosmological hydrodynamics 
simulation of galaxy formation which includes cooling \cite{ref9, ref10}, 
regardless of the technique employed to solve the Euler equations. 
Such an oversight is predicted
to have drastic consequences on the ability of these models to match star
formation rates and supernova driven gas outflows. Arguably, the most
straightforward observational tracer of the cold flow scenario
is the UV flux emitted by galaxies, as it mainly
originates from newly born stars. However, the matter is rapidly
complicated by the presence of dust, which is generated on supernovae
timescales ($\approx$ 50 million years) and is very efficient at absorbing UV
radiation. The spatial distribution of dust grains relative to stars is
therefore a crucial ingredient in any estimate of the UV luminosity
function and it is notoriously difficult to predict (semi-)
analytically \cite{ref11, ref12}. Accordingly, we address this
key issue numerically.}

\section{Setup and extinction model}

{
The evolution of a cubic cosmological volume of 50h$^{-1}$ Mpc on a side (comoving), containing
1024$^3$ dark matter particles and an Eulerian root grid
of 1024$^3$ gas cells is followed. A $\Lambda$CDM
concordance universe ($\Omega_{m} = 0.3, \Omega_{L} =
0.7, h=H_{0}/[100 \ {\rm km} s^{-1}
{\rm Mpc}\textsuperscript{-1}] =0.7, \sigma_{8} = 0.9, n = 1$, i.e.
the WMAP 1 year best fit cosmology \cite{ref13}, is adopted,
resulting in a dark matter particle mass $m_{p} = 1.41\times
10^{7} M_{{\sun}}$ and we use an
adaptive mesh refinement (AMR) technique \cite{ref14} to keep our
spatial resolution fixed at around 1 h$^{-1}$kpc in
physical coordinates. }

{ The matter density fields are evolved from z =
120 to z = 4, and we output about 34 snapshots regularly spaced in the
expansion factor. In each snapshot, we identify dark matter halos as
well as the subhalos they contain using the {\sc AdaptaHOP}
algorithm \cite{ref15}, and only keep groups with more than 100
particles. We measure the basic physical properties of these groups and
subgroups, such as their masses M$_{\rm hop}$ and radii
$R_{\rm hop}$. Key physical processes for galaxy formation such
as radiative cooling, UV background radiation, star formation and
supernovae feedback are also implemented self-consistently in
{\sc ramses} \cite{ref36,ref17}. Gas cools down to a minimum of 10$^4$ K at a rate which depends on
the local metallicity in each grid cell, and turns into stars on a
constant timescale of $t_\star = 2 $Gyr. Stars then evolve,
releasing an amount of metals and energy into the ISM specified by a
Salpeter initial mass function and the physics of a Sedov
explosion \cite{ref17}. In some cases, the combined
supernovae blow part of the ISM back into the halo's
hot phase and further into the IGM. For each time snapshot of the
simulation we associate the stars located within each dark matter
substructure detected by {\sc AdaptaHOP} to a different galaxy. Note
that, proceeding this way, we account for \textit{all} the stars
created in the simulation. By keeping track of the stellar content of
each galaxy, as a function of age and metallicity, and knowing the
galaxies' gas content and chemical composition, we then compute the
(possibly extinct) spectral energy distribution of each galaxy, along
the lines described in the {\sc stardust} model \cite{ref11}.}

{
More specifically, we calculate the spectrophotometric properties of galaxies including
internal extinction of starlight by dust. We model dust absorption 
within each galaxy using the empirical calibration of
Guiderdoni \& Rocca-Volmerange (1987), in which the optical
depth scales with metallicity and in proportion to the column density
of hydrogen along the line of sight. The optical depth, $\tau$, is therefore given by: }
{
\begin{equation}
	\tau(\lambda) = \left(\frac{A_\lambda}{A_{\rm V}}\right)_{{\rm Z}_{\sun}}
		\left(\frac{Z_{\rm g}}{{\rm Z}_{\sun}}\right)^s
			\left(\frac{N_{\rm H}}{2.1 \times 10^{21}\;{\rm cm}^{-2}}\right),
\end{equation}
where $\lambda$ is the wavelength, and $(A_\lambda/A_{\rm V})_{{\rm Z}_{\sun}}$  is the extinction curve for solar
metallicity, which we take to be that of the Milky Way \cite{ref19}.
We adopt a scaling with metallicity of s=1.6 for ${\lambda}$ ${\geq}$
2000{\AA} and s=1.35 for ${\lambda}$ ${\leq}$ 2000{\AA} \cite{ref18}.
We calculate the extinction for subsamples of galaxies in logarithmic bins of stellar mass (100
objects per bin), based on the actual distribution of gas and metals in
each galaxy and averaging over lines of sight. The resulting
distributions are then used to draw random extinctions for galaxies
with similar stellar masses. The total number of galaxies varies
between 43000 and 86000 between z=7 and z=4.}
\section{The UV LF of high-z galaxies}
{
Observationally, the largest samples of UV emitting galaxies at
$z > 3$ are selected according to a
colour-colour criterion that maps the Lyman break feature in their
spectral energy distributions (hence their name ``Lyman-Break
Galaxies'' or LBG \cite{ref20}). This now well-established technique
allows the detection of large numbers of galaxies per field of view
using relatively little observing time, and  does not strongly bias the selection process
\cite{ref22,ref23,ref24}. 
Such data sets provide us
with the best sample of high-z galaxies against which to perform
a meaningful statistical comparison. }
{
The most robust and simple statistics to consider for such a comparison
is the LF of galaxies. This is the
main focus of this letter. 

Like the observations, which are limited by
the resolving power of the telescope/instrument combination with which
they are obtained, the LFs measured in numerical simulations suffer
from finite box size and resolution effects, so their domain of
validity is restricted to a certain range of magnitudes. In the
simulation presented here (Fig 1), dark matter halos smaller than
$\approx 10^{9} M_{\sun}$ 
in mass are barely resolved in terms of the hydrodynamics of their gas
content, so galaxies hosted by smaller halos are absent, which
introduces a non-physical cutoff at the faint end of the LF (Fig.
3). As spectral energy distributions of galaxies are computed as the
weighted sum of single stellar populations, we avoid being dominated by
Poisson noise by restricting our results to galaxies containing at
least 10 star particles. At the bright end, objects become rarer and
rarer, until we reach the point where we run out of bright galaxies.
This effect is somewhat amplified by our finite simulated volume, which
is small compared to the size of the current observable Universe: at
some point, the statistical error on the LF becomes of the order of the
measurement itself. Vertical and horizontal dashed lines on Fig. 3 mark
these numerical limitations respectively, clearly delineating the
region where our LFs can be trusted. Note that, although approximate, 
these are conservative boundaries. A truly accurate determination would require running several simulations
encompassing larger volumes and featuring a higher mass resolution.
Finally, despite our resolution, still short of being able to provide a
fully self-consistent model for re-ionization by about an order of
magnitude in mass \cite{ref25,ref26}, the UV background radiation has
little effect on the galaxies discussed here \cite{ref27} since the
minimal circular velocity of their host dark matter halo is larger than
25 km/s \cite{ref28}.

Setting aside the star formation rate and stellar
initial mass function, the key issue to derive correct UV LFs lies in
our ability to estimate dust extinction accurately, since dust grains
very easily absorb light in this wavelength range. Because a full
modelling of the formation and evolution of the dust grain population
is beyond the reach of current simulations, we have tried a couple of
phenomenological models to assess the impact of extinction on our
results. First, as a sanity check, we use the most widespread method in
the literature, which assumes a universal shape for the extinction
curve \cite{ref29}, and then fits its normalization,
E(B$-$V), to obtain the best possible match to the observational
luminosity function \cite{ref30}. This means that a unique ``average'' extinction is
assigned to every galaxy for which we compute a UV magnitude. We
believe this is, at best, a very awkward state of affairs, since it is
empirically known that, at least in the local Universe, the value of
the extinction is correlated to the UV magnitude, such that brighter
galaxies have larger values of E(B $-$ V) \cite{ref4}. However, as pointed out by 
Night et al. (2006), invoking luminosity dependent extinction can be
seen as a cheat, as it is equivalent to hiding the discrepancy between
simulated and observed LFs in the extinction law. As a more ambitious
alternative, we therefore measure the distribution of gas and metals on
the line of sight towards each of the star particles in our simulated
galaxies, which allows us to calculate the optical depth as a function
of wavelength for each of these lines of sight (see Fig. 2). The main
caveat of this second approach, is the assumption that dust properties
only depend on these two quantities (at least for absorption of UV
light), but it was shown \cite{ref18} that this seems to be a fair description
for extinction in local galaxies. Arguably the most important feature of such a dust model is that it
provides a self-consistent method to compute the extinction law
directly from the simulation itself.}

\begin{figure}
\center\includegraphics[angle=90,width=8.5cm,angle=0]{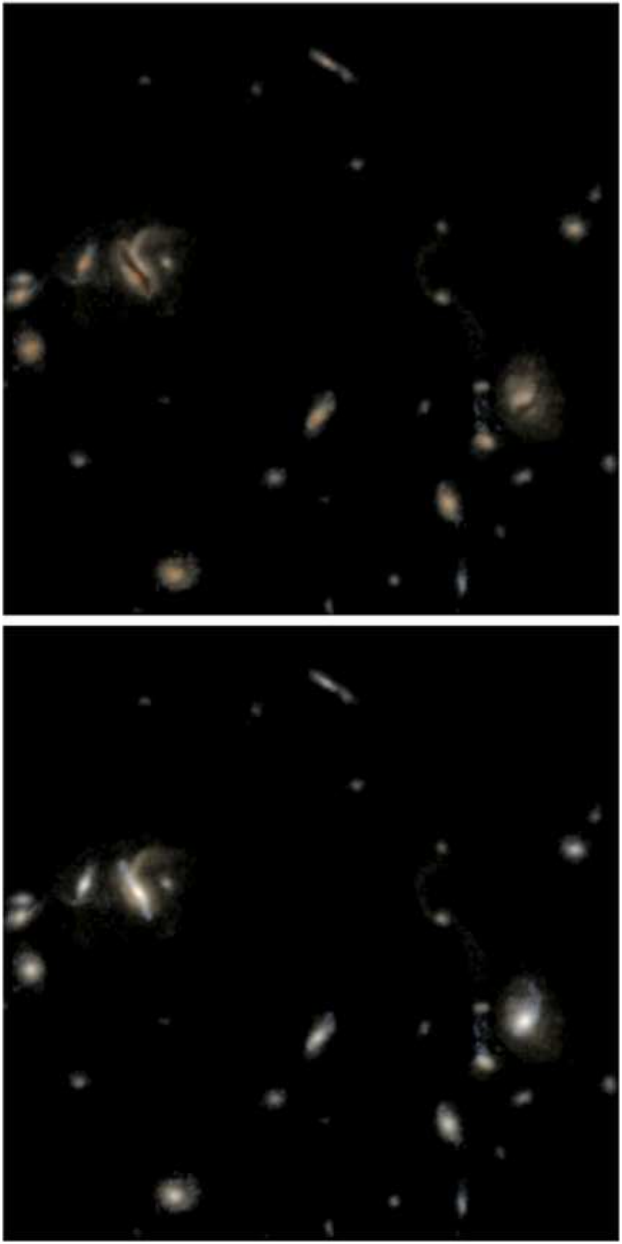}
\caption{
 A typical field of view of 1 comoving Mpc on
a side in the Mare Nostrum simulation shown in composite rest frame
true colours (R, B, and V filters). Colours of the stellar population
are shown in the left panel without dust extinction, and with
extinction in the right panel. Note the presence of dust lanes
in the two largest galaxies, testimony of the highly non-homogeneous
distribution of dust and stars in the ISM.}
\end{figure}

{The two dust models produce LFs
that match the observed one quite well within the whole range of
magnitudes it covers, as observational error bars plotted on fig. 3
only roughly estimate cosmic variance errors \cite{ref31}. We discuss 
our results in terms of the most
commonly used analytical parameterisation of the galaxy LF, i.e. the
Schechter function, examples of which are plotted on fig.
3. This description relies on three parameters, $\Phi_\star$, M$_\star$ and $\alpha$, i.e. the
normalisation, characteristic magnitude and faint-end slope of the
LF, respectively. The preferred Schechter fit for our
LF at z=4 (i.e. the green curve on the bottom right
panel of fig. 3) yields a faint end slope of 1.6 and a
characteristic magnitude of 20.9, compared to 1.73+/0.05,
20.98+/0.1 for the data respectively
\cite{ref33}. Our normalisation of 0.003 Mpc$^3$
appears a bit high with respect to that measured in the data
(0.0013+/0.0002 Mpc$^3$). However, this is the
parameter of the Schechter function which is the most sensitive to a
change in cosmological model. In other words, the error we make when we
rescale our LFs from the Wilkinson Microwave Anisotropies Probe (WMAP)
1 year \cite{ref13} to the WMAP 5 year parameters (green curves on
fig. 3, $\Omega_{m} = 0.26, \Omega_{\Lambda} = 0.74,
h=H_{0}/[100 km s^{-1}
Mpc^{-1}] =0.72, \sigma_{8} = 0.8, n = 0.96$)
\cite{ref34} mostly affects $\Phi_\star$, so that the factor of about two
difference is not statistically discrepant. To be more precise, in order to recast the
results of our simulation for a different set of cosmological
parameters, we have used the Press \& Schechter (1974) \nocite{ref32} formalism
to calculate the dark matter halo mass function at a given z in
both cosmologies. Assuming an occupation number of one galaxy per dark
matter halo, we then divide our luminosity functions by the ratio of
the dark matter halo mass functions. Given the mass resolution of our
simulation, we expect this correction to be quite accurate at higher
redshift (above z=6) and to slowly degrade as more massive haloes
begin to form and start hosting a larger number of galaxies on average.
This explains, at least partially, the larger discrepancy (at the $\sim$ 2
sigma level once cosmic variance is taken into account) between model
and observed LFs at intermediate and low luminosities for
z $<$ 6. Note that these considerations do not apply to the
brightest objects, which are much more scarce and have a much higher
probability of being unique in their host dark matter halo. As a
result, the correction remains valid for these objects even for
z{\textless}6, as can be seen on fig. 3.

\begin{figure*}
  \includegraphics[angle=0,width=14cm]{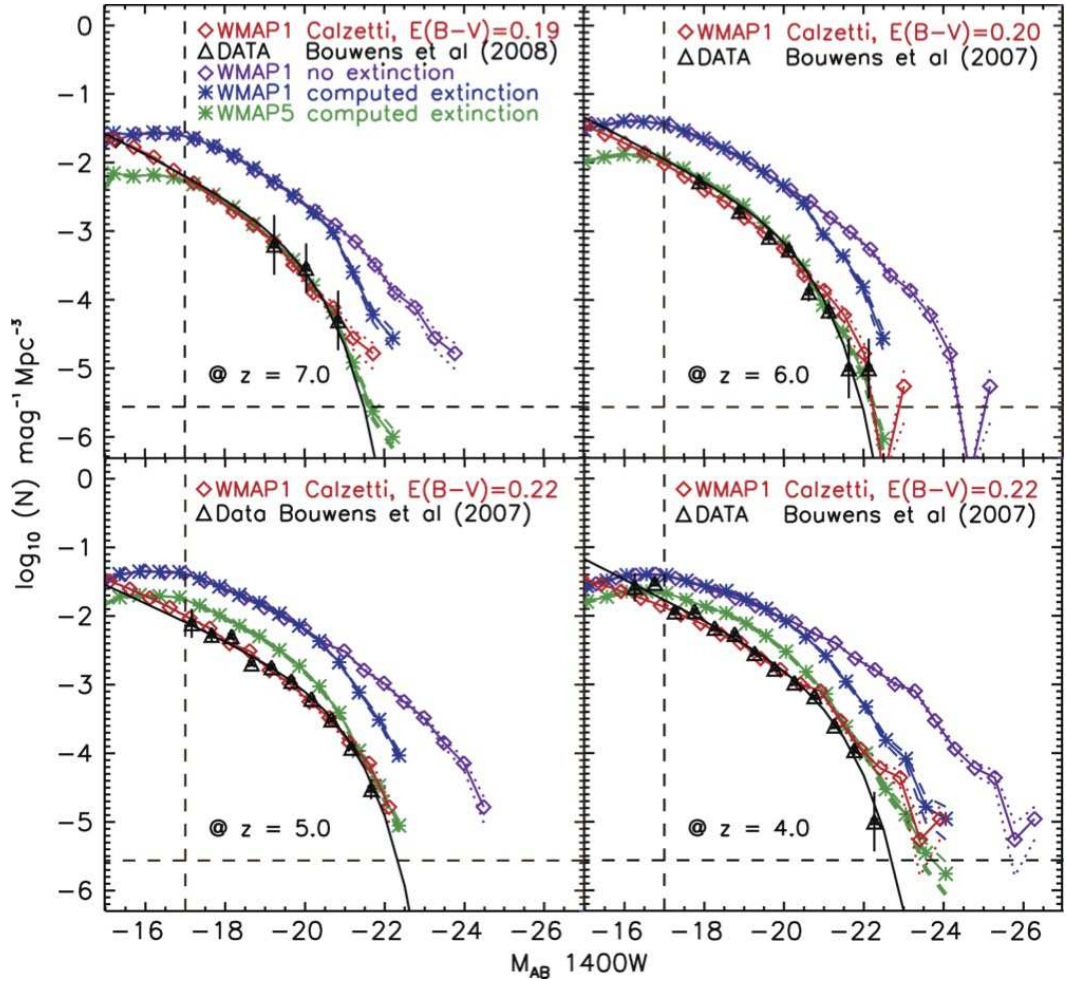}
\caption{
Redshift evolution of the UV (1400 {\AA}) rest frame
galaxy luminosity function measured in the Mare Nostrum simulation.
Vertical dashed lines indicate mass resolution, horizontal dashed lines
volume resolution. Purple diamonds
indicate nonextinguished WMAP-1 cosmology
LFs, blue stars LFs extinguished with a
dust content scaled using the measured abundance of metals and gas in
the simulation. Red diamond LFs stand for galaxies uniformly
extinguished with a 
\protect\cite{ref29}
law in a WMAP-1 universe, and
green stars for LFs extinguished as the blue star ones but rescaled to
the currently favoured WMAP-5 cosmology. 
Black triangles and solid lines mark
data gathered by Bouwens and collaborators
 \protect\cite{ref31,ref33}
}
\end{figure*}

{
In a nutshell, we find that our
simulated LFs at z$\sim$4 are in respectable agreement with the
available UV data and robustly predict a faint-end slope with a value
${\alpha}\,{\approx}-1.6$ compatible with the findings of Steidel
and collaborators \cite{ref20,ref35} (${\alpha}$ ${\approx}$ $-$1.6) but
not with the slope found in the Subaru data \cite{ref16} (${\alpha}$ ${\approx}$ $-$2.2).
Our LFs also show quite an important amount of evolution, as measured by M$_\star$,
which varies from 19.8 at z=7 to 20.9 at z=4. This strong
evolution is a direct consequence of the rapid decrease of the cosmic
star formation rate density from z = 4 to 7 in our simulation and is,
to a large extent, unaffected by resolution effects \cite{ref36}. At z$>$6, 
recent analyses of the Hubble Space Telescope Ultra
Deep Field and the Great Observatories Origins Deep Survey data
\cite{ref33,ref31} suggest a slightly steeper faintend slope of
${\alpha}$ ${\approx}$ $-$1.74, again in good agreement with our
simulation as we find that the faint-end slope of our LF does steepen
slowly with redshift. This is of key importance since ionising photons
from low luminosity galaxies could possibly suffice to carry out the
reionisation of the Universe, provided these latter are numerous
enough. Reversing the argument and marching down in redshift, this
raises the question of what mechanism is responsible for such a
flattening of the faint end slope of the UV LF, which extends down to the local
Universe, since a shallow slope (${\alpha}$ ${\approx}$
1.22) is observed locally for z $<$ 0.1 in the
 GALEX  far-UV data
  \cite{ref37}. Supernova feedback
could potentially drive powerful galactic outflows and quench star formation, but 
SPH
simulations that explicitly include winds
efficient enough to drive gas out of low mass halos, systematically
measure a steeper faint end slope than we do (${\alpha}$ ${\approx}$
$-$2.0) \cite{ref30}. Finally, at the bright end of the LF, our two
dust extinction models yield different results. The self-consistent
extinction model has a more pronounced cut-off, which better matches
the shape of a Schechter function. This feature stems from an
intrinsically stronger dust attenuation that occurs in the most UV
bright galaxies. We therefore predict that these very same UV beacons
also are the most luminous IR sources in the sky. The forthcoming
generation of sub-millimetre facilities, and in particular ALMA,
will either validate or disprove this prediction.}

\section{
Conclusions and discussion}

{
We have derived UV LFs of simulated high-z
galaxies using the largest hydrodynamical cosmological simulation
performed so far. Including a self-consistent post-treatment of the
extinction in our galaxies based on the abundance of gas and metals
measured in the simulation, we have compared these synthetic LFs to the
observed ones. We have studied how sensitive these LFs are to both
cosmological parameters and dust absorption, using the most
commonly used extinction model in the community \cite{ref29} as a
reference. Our self-consistent treatment of dust absorption enables
us to match the observed UV LFs. This match quantitatively compares to
the fit obtained with the simple model where all the galaxies are
dimmed in the same way, provided that we set the cosmological
parameters (in particular the normalization of the power spectrum and
its rolling index) to their best fit WMAP 5 years values \cite{ref34}.
However, contrarily to the Calzetti extinction model, such an agreement
is achieved without the introduction of an extra free parameter, i.e.
the average extinction of the galaxy population as a function of
redshift can be retrieved from the simulation itself. Moreover, this
selfconsistent model receives support from (i) the extrapolation of
low z observations which show that extinction depends on UV
luminosity \cite{ref4} (ii) the mounting evidence for an extremely low
metallicity in faint z$\sim$7 galaxies \cite{ref38}, which both favour a
non-uniform degree of extinction throughout the magnitude range
spanned by the LF.}

{
Although current observational surveys still probe small volumes of the
high redshift universe, and are therefore prone to suffer from cosmic
variance at the bright end of the LF, they seem to go deep enough to
yield reliable estimates of its faint end slope in the UV, up to
z$\sim$7. At fixed cosmological parameters, the key to
matching UV luminosities with any model of galaxy formation is the
amount of dust extinction present in galaxies. However, the absorbed UV
light is reprocessed by dust, and astronomers should be able to
detect the far infrared counterparts of low luminosity primordial
galaxies, provided extinction strongly affects galaxies across their
entire luminosity range. Future wider surveys together with the advent
of infrared and deep millimetre observations from the JWST and  ALMA respectively should therefore give us a robust
handle on where energy from star formation is really coming out at
these redshifts. These measurements will, in turn, confirm or
invalidate our improved treatment of dust attenuation. Indeed, we
predict that this attenuation should strongly correlate with UV
luminosity, with low luminosity galaxies almost dust free at z$\sim$7,
and massive galaxies extinguished by 1.5 to 2 magnitudes. Hence, only
the tip of the iceberg of primordial galaxy formation should be visible
at rest-frame infrared wavelengths if our model is correct: the bulk
of the galaxy population will mostly be UV bright with a few
isolated monsters shining across the entire wavelength range.}

\eject
\noindent{\bf Acknowledgments \\}
This work was supported by ANR grant number 05441478 awarded to
the Horizon Project. CP acknowledges support from a Leverhulme visiting
professorship at the University of Oxford. The simulation was run on
the Mare Nostrum machine at the BSC and we
thank the staff for their competent support and warm
hospitality.
\bibliographystyle{mn2e}

\end{document}